\pdfoutput=1
\documentclass[12pt]{article}
\usepackage{epsfig, epsf, graphicx}
\usepackage{pstricks, pst-node, psfrag}
\usepackage{amssymb,amsmath,bm}
\usepackage{verbatim,enumerate}
\usepackage{rotating, lscape}
\usepackage{setspace}
\usepackage{multirow}
\usepackage{float}
\usepackage{wrapfig}
\usepackage{caption}
\usepackage{subcaption}

\usepackage[round]{natbib}
\def\boxit#1{\vbox{\hrule\hbox{\vrule\kern6pt
          \vbox{\kern6pt#1\kern6pt}\kern6pt\vrule}\hrule}}

\def\bse{\begin{eqnarray*}}
\def\ese{\end{eqnarray*}}
\def\be{\begin{eqnarray}}
\def\ee{\end{eqnarray}}
\def\bq{\begin{equation}}
\def\eq{\end{equation}}
\def\bse{\begin{eqnarray*}}
\def\ese{\end{eqnarray*}}

\def\trans{^{\rm T}}

\newcommand{\btheta}{\boldsymbol{\theta}}

\begin{document}

\thispagestyle{empty} \baselineskip=28pt \vskip 5mm
\begin{center} {\Huge{\bf Non-Gaussian Autoregressive Processes with Tukey $g$-and-$h$ Transformations}}
\end{center}

\baselineskip=12pt \vskip 10mm

\begin{center}\large
Yuan Yan$^1$ and Marc G.~Genton\footnote[1]{
\baselineskip=10pt Statistics Program,
King Abdullah University of Science and Technology,
Thuwal 23955-6900, Saudi Arabia.\\
E-mail: yuan.yan@kaust.edu.sa, marc.genton@kaust.edu.sa\\
This publication is based upon work supported by the King Abdullah University of Science and Technology
(KAUST) Office of Sponsored Research (OSR) under Award No: OSR-2015-CRG4-2640.}
\end{center}

\baselineskip=17pt \vskip 10mm \centerline{\today} \vskip 15mm

\begin{center}
{\large{\bf Abstract}}
\end{center}
When performing a time series analysis of continuous data, for example from climate or environmental problems, the assumption that the process is Gaussian is often violated. Therefore, we introduce two non-Gaussian autoregressive time series models that are able to fit skewed and heavy-tailed time series data. Our two models are based on the Tukey $g$-and-$h$ transformation. We discuss parameter estimation, order selection, and forecasting procedures for our models and examine their performances in a simulation study. We demonstrate the usefulness of our models by applying them to two sets of wind speed data.
\baselineskip=14pt

\par\vfill\noindent
{\bf Some key words:} Autoregressive; Heavy tails; Non-Gaussian; Skewness; Time series; Tukey $g$-and-$h$ transformation.
\par\medskip\noindent
{\bf Short title}: Tukey $g$-and-$h$ Autoregressive Processes

\clearpage\pagebreak\newpage \pagenumbering{arabic}
\baselineskip=26pt

\section{Introduction}\label{sec:intro}
To study climate change, it is essential to understand the temporal properties of the data of interest. Climate and environmental data, such as precipitation, temperature, thickness of glacial varves, wind speed, concentration of air pollutants, are continuous in nature. In a time series analysis of continuous random variables, the autoregressive moving-average (ARMA) model, especially the Gaussian one, is popularly adopted because of its simplicity and interpretability. However, data from climate or environmental science are often asymmetric with heavy tails. Therefore, a non-Gaussian time series analysis is needed. Since most climate or environmental data we come across are continuous, in this paper we do not consider time series analysis of categorical or binary data, which are intrinsically non-Gaussian. 

In linear models, three approaches are mainly used to deal with non-Gaussian data: transformation of the dependent variable, regression with non-Gaussian error, and generalized linear models (GLM). Similarly, we can also classify many of the existing non-Gaussian time series methods and models into three categories: 1) transformations to `Gaussianize' the data, 2) ARMA models with non-Gaussian noise, and 3) extension of the GLM to the time series context.

The transformation approach is widely used among practitioners to, first, Gaussianize data and then examine the latent Gaussian process. Equivalently, it assumes that the observed process $Y_t$ is obtained after a transformation $\eta$ of the latent Gaussian process $Z_t$:
\begin{equation}
Y_t=\eta(Z_t).\label{model1}
\end{equation}
For example, precipitation data can be approximated by the gamma distribution, then a square or cube root transformation is usually applied to normalize the data; similarly wind speed data can be assumed to follow a Weibull distribution, and taking a logarithm alleviates the departure from normality \citep{Has89}. The logarithm, square and cube root transformations all belong to the Box-Cox power transformation \citep{BoxCox64}, which is a family of transformations with one parameter ($\lambda$) and can be applied only to positive values. A parametric form of transformations allows users to find the optimal transformation $\eta$ from among the members of the Box-Cox transformation family by estimating $\lambda$ from the data, instead of attempting many different transformations. 
A large literature exists on the Box-Cox power transformation, as well as its modifications and its application in linear models and time series analysis. \citet{Nel79} extensively explored the use of Box-Cox transformations in 21 time series from economics. Non-parametric transformations are an appealing alternative to assuming a parametric form of transformations due to their flexibility. \citet{Nychka03} explored non-parametric transformations on both the dependent variable and the predictor variables in regression. \citet{Block90} used the empirical distribution and probability integral transform (PIT) to form ARMA models with arbitrary continuous marginal distributions. However, there is a trade-off between flexibility and parsimony. 

Instead of using transformations to Gaussianize the data, an ARMA model with non-Gaussian noise may be specified, whether it is the marginal distribution of the process or the distribution of the error term. Given the marginal distribution, the distribution of the error term can be found by linking their moment generating function, namely, the Laplace transformation of the probability density function. Many models were previously proposed following this direction, for example, ARMA models with exponential marginals \citep{EARMA} and AR models with marginal gamma distribution \citep{GARMA}. On the other hand, specifying the distribution of the non-Gaussian error term is easier for data simulation and maximum likelihood inference since the conditional likelihood can then be written down in a closed form. ARMA models that are driven by non-Gaussian innovations and their properties were discussed in \citet{Da80} and \citet{Li88}. 

For the third class of existing non-Gaussian time series methods and models, \citet{Ben03} introduced the generalized ARMA model, extending the GLM to a time series setting. Earlier exploration of GLM in time series can be found in \citet{Ben03} and references therein. \citet{Cor09} proposed a model combining the GLM idea with transformations, and considered its use in time series. 
 
Apart from the three categories summarized above, there exists many other specialized models for non-Gaussian, non-stationary time series. The most famous ones are the autoregressive conditional heteroskedasticity (ARCH) \citep{ARCH} and the generalized autoregressive conditional heteroskedasticity (GARCH) \citep{GARCH} models. These models have been used canonically for analyzing financial time series that exhibit changes in volatility. Other models include the zeroth-order self-exciting threshold autoregressive (SETAR) model \citep{tong90}, the multipredictor autoregressive time series (MATS) models \citep{martin92}, the Gaussian mixture transition distribution (GMTD) models \citep{Le96}, the mixture autoregressive models \citep{Wo00}, and their extension to the Student $t$-mixture autoregressive model \citep{Wo09}.

Considering the trade-offs between model flexibility and parsimony, in this paper we use an alternative parametric family of transformations, the Tukey $g$-and-$h$ (TGH) transformation, which is more flexible and interpretable than the Box-Cox family and at the same time keeps the model fairly simple compared to the non-parametric approaches. We build two autoregressive models for non-Gaussian time series via the TGH transformation in both the data transformation and non-Gaussian error approaches. Our first model uses the TGH transformation for $\eta$ in (\ref{model1}). Our second model assumes the non-Gaussian error term in the AR process to be a TGH transformation of Gaussian white noise. 
 
The remainder of this paper is organized as follows. In Section~\ref{sec:model}, we review the TGH transformation, discuss properties of the univariate TGH distribution and its extensions, and introduce our two AR models based on the TGH transformation. In Section~\ref{sec:est_pred}, we describe the parameter estimation and forecasting algorithm for our models. In Section~\ref{sec:sim}, we demonstrate the estimation and prediction performances of our models through a simulation study. In Section~\ref{sec:app}, we illustrate the usefulness of our models by applying them to two wind speed datasets. We summarize our findings and prospects for future research in Section~\ref{sec:sum}.
\section{Two Tukey $g$-and-$h$ Autoregressive Models}\label{sec:model}
\subsection{TGH Transformations and Properties}\label{sec:prop}
The TGH transformation \citep{tukey77} is defined as a strictly monotonic increasing function of $z\in\mathbb{R}$ for $h\geq 0$ and $g \in \Bbb{R}$:
\begin{equation}
\tau_{g,h}(z)=\begin{cases}g^{-1}\{\exp(gz)-1\}\exp(hz^2/2),&g\neq 0,\\z\exp(hz^2/2),&g=0.\end{cases} \label{taugh}
\end{equation}
When applying the TGH transformation to a univariate standard normal random variable $Z \sim \mathcal{N}(0,1)$, the transformed random variable $T=\tau_{g,h}(Z)$ is said to follow the TGH distribution. The support of $T$ is the real line when $h\neq 0 \text{ or } h=g=0$; for $h=0$ and $g \neq 0$, it has a lower (upper) bound of $-1/g$ when $g>0$ ($g<0$). The univariate TGH distribution is a flexible family of distributional models for non-Gaussian data. It has two interpretable parameters, $g$ and $h$, which respectively control the skewness and tail heaviness of $Y$. The tails become heavier as $h$ increases, and the $q$-th moment of $Y$ exists only for $h<1/q$. The level of skewness increases as $|g|$ increases (becoming right-skewed when $g>0$). \citet{Martinez84} included the case of $h<0$, for which the TGH transformation is no longer monotonic and the resulting TGH distribution has lighter tails than the Gaussian one. However, their method is unconventional, so we only consider the case of $h\geq0$. Also, from a practical point of view, real data are more often heavy-tailed rather than light-tailed. Special cases of the TGH distribution include the shifted log-normal random variable for $h=0$ and $g>0$, and a Pareto-like distribution for $g=0$ and $h>0$, which was discussed in detail by \citet{Georg15}. The $q$-th moments of the TGH distribution were derived by \citet{Martinez84}, for $h<1/q$:
\begin{equation}
\mathbb{E}(T^q)=\begin{cases}\frac{1}{g^q\sqrt{1-qh}}\sum\limits_{i=0}^{q}(-1)^i{q \choose i}\exp\left\{\frac{(q-i)^2g^2}{2(1-qh)}\right\},&g\neq 0,\\
\frac{q!}{2^{q/2} (q/2)!}(1-qh)^{-\frac{q+1}{2}}, &g=0, q \text{ even},\\
0, &g=0, q \text{ odd}. \end{cases} \label{momen}
\end{equation}
We illustrate various aspects of the Tukey $g$-and-$h$ distributions by a visuanimation \citep{Visu} in the supplementary material.
  

Extensions of the univariate TGH distribution to the multivariate case can be found in \citet{FG06} and \citet{HR12}. \citet{XG16} used the TGH transformation to build a new max-stable process for the modeling of spatial extremes. Recently, \citet{XG17} further extended the TGH distribution to the spatial case and defined TGH random fields. Those models were found to be useful in practice and have been applied to air pollution, wind speed, rainfall and economic data. In this paper, we apply the TGH transformation in the time series context and build two models that take advantage of the AR structure. 

\subsection{TGH Transformation of a Latent AR Process}\label{sec:model1}
Our first model transforms a latent Gaussian process similarly to \citet{XG17}. Let $Z_t \sim \text{AR}(p),\ t=1,2,\ldots$, be a stationary Gaussian AR process of order $p$ with zero mean and unit variance, and let $T_t=\tau_{g,h}(Z_t)$ by applying the standard TGH transformation to $Z_t$. Our model 1, the Tukey $g$-and-$h$ transformed autoregressive (TGH-AR($p$)-t) process $Y_t$, has the same form as (\ref{model1}), with $\eta$ being a generalized version of the TGH transformation: 
\begin{equation}
Y_t = \bm{X}_t\trans\bm{\beta}+\xi+\omega\tau_{g,h}(Z_t), \label{model11}\end{equation}
where $\bm{X}_t$ and $\bm{\beta}$ are the covariates and the corresponding coefficients respectively, and $\xi$ and $\omega$ are the location and scale parameters, respectively. In a time series analysis, the covariates are usually functions of $t$ to model the trend or the seasonality, which may also be incorporated through a seasonal integrated AR model. In our first model, unlike the commonly adopted Box-Cox approach, the deterministic part that consists of the location parameter, covariates and their coefficients, is outside the transformation. In this way, the trend (also seasonality or periodicity) is attributed directly to the process $Y_t$ rather than the underlying Gaussian process $Z_t$. We control the skewness and tail behavior of the stochastic part of the process $Y_t$ by applying the TGH transformation to $Z_t$ with different values of $g$ and $h$.  
The moments of $Y_t$ can be computed as the moments for $T_t$ in (\ref{momen}) with modification for the additional location and scale parameters. The autocovariance function $C_T(\tau)$ of $T_t$ can be found in terms of the autocorrelation function (ACF) $\rho_Z(\tau)$ of the underlying Gaussian process $Z_t$, which was derived in \citet{XG17}:
\begin{align*} 
C_T(\tau)&=\frac{\exp\left[\frac{1+\rho_Z(\tau)}{1-h\{1+\rho_Z(\tau)\}}g^2\right]-2\exp\left[\frac{1-h\{1-\rho_Z^2(\tau)\}}{(1-h)^2-h^2\rho_Z^2(\tau)}\frac{g^2}{2}\right]+1}{g^2\sqrt{(1-h)^2-\rho_Z^2(\tau)h^2}}-\left\{\mathbb{E}({T_t})\right\}^2.
\end{align*}
The autocorrelation of $Y_t$ is always weaker than the autocorrelation of $Z_t$ and in the supplementary material we illustrate this fact via Visuanimation~S1. 

When $p=1$, the TGH-AR(1)-t process is of the form (\ref{model11}), with $Z_t=\phi Z_{t-1}+\epsilon_t$, and where $\epsilon_t$ is a Gaussian white-noise process with zero mean and constant variance $\sigma^2_\epsilon=1-\phi^2$. There is a constraint that $\sigma^2_\epsilon$ is a function of $\phi$ so that $Z_t$ has unit variance. This dependency may seem strange at first, but it is effectively equivalent to standardizing $Z_t$ to a process with unit variance by multiplication by a scaling constant. For example, the usual AR(1) process with a Gaussian error term $\epsilon_t$ of variance $\sigma_{\omega}^2$ will result in a process $Z_t^\prime$ of variance $\frac{\sigma_{\omega}^2}{1-\phi^2}$, and $Z_t^\prime$ can be standardized to have a unit variance by multiplying the rescaling factor by $\frac{\sqrt{1-\phi^2}}{\sigma_{\omega}}$. This is equivalent to an AR(1) process with an error term of variance of $1-\phi^2$. 

\begin{figure}[h!] 
	\centering
	\includegraphics[width=\textwidth]{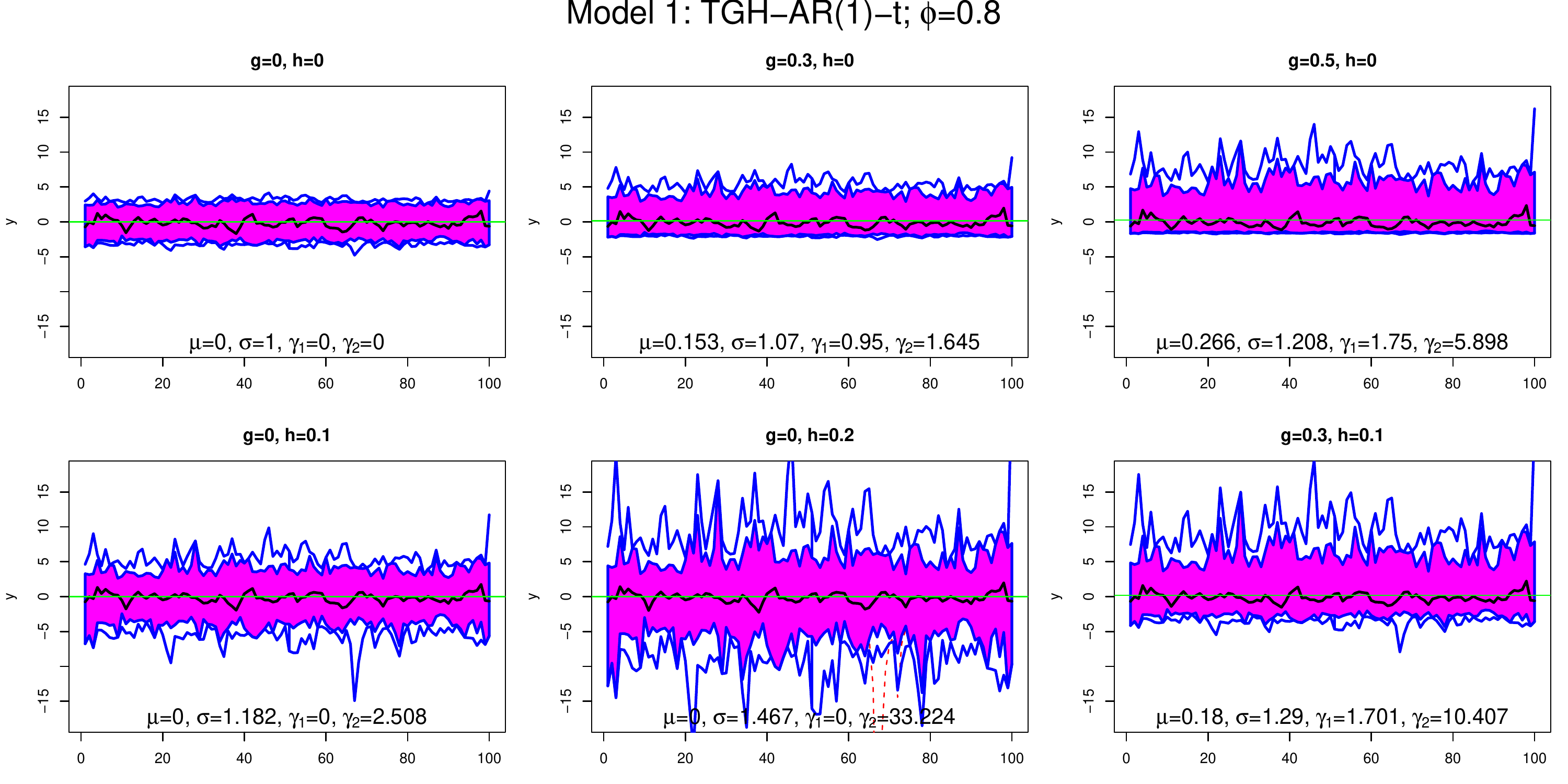}	\vspace{-3mm}		
	\caption{Functional boxplots of 1000 realizations of sample size $n=100$ from model 1, i.e., TGH-AR(1)-t, without covariates, with $\xi$=0, $\omega$=1,  an AR coefficient of $\phi$=0.8 and six pairs of values for $g$ and $h$, labeled with values of the mean ($\mu$, green line), standard deviation ($\sigma$), skewness ($\gamma_1$) and excess kurtosis ($\gamma_2$) of the process.}
	\label{fb1}
\end{figure}

To simulate samples from the TGH-AR($p$)-t model, we first simulate data from the underlying Gaussian AR($p$) process $Z_t$ for $t=1,2,\ldots,n$. Then we transform the process by applying the generalized version of the TGH transformation to $Z_t$ at each time point. Figure~\ref{fb1} shows the functional boxplots \citep{SG11} of 1000 realizations from the TGH-AR(1)-t model of sample size $n=100$ without covariates, with $\xi=0$, $\omega=1, \phi=0.8$, $\sigma^2_\epsilon=0.36$ and six pairs of values for $g$ and $h$. The six pairs of values for $g$ and $h$ include the Gaussian AR case when $g=h=0$. 
The functional boxplots are labeled with values of the marginal mean ($\mu$, green line), standard deviation ($\sigma$), skewness ($\gamma_1$) and excess kurtosis ($\gamma_2$) for the transformed process $Y_t$. 
From the functional boxplots of the sample paths of $Y_t$, we see clearly when $h=0$, as $g$ increases from 0.3 to 0.5, the process is more right skewed;  when $g=0$, as $h$ increases from 0.1 to 0.2, extreme values occur more in the time series; with $g=0.3, h=0.1$ we see the effect of both $g$ and $h$ on the process.

\subsection{AR Process with TGH Error}\label{sec:model2}
\begin{figure}[th!] 
	\centering
	\includegraphics[width=\textwidth]{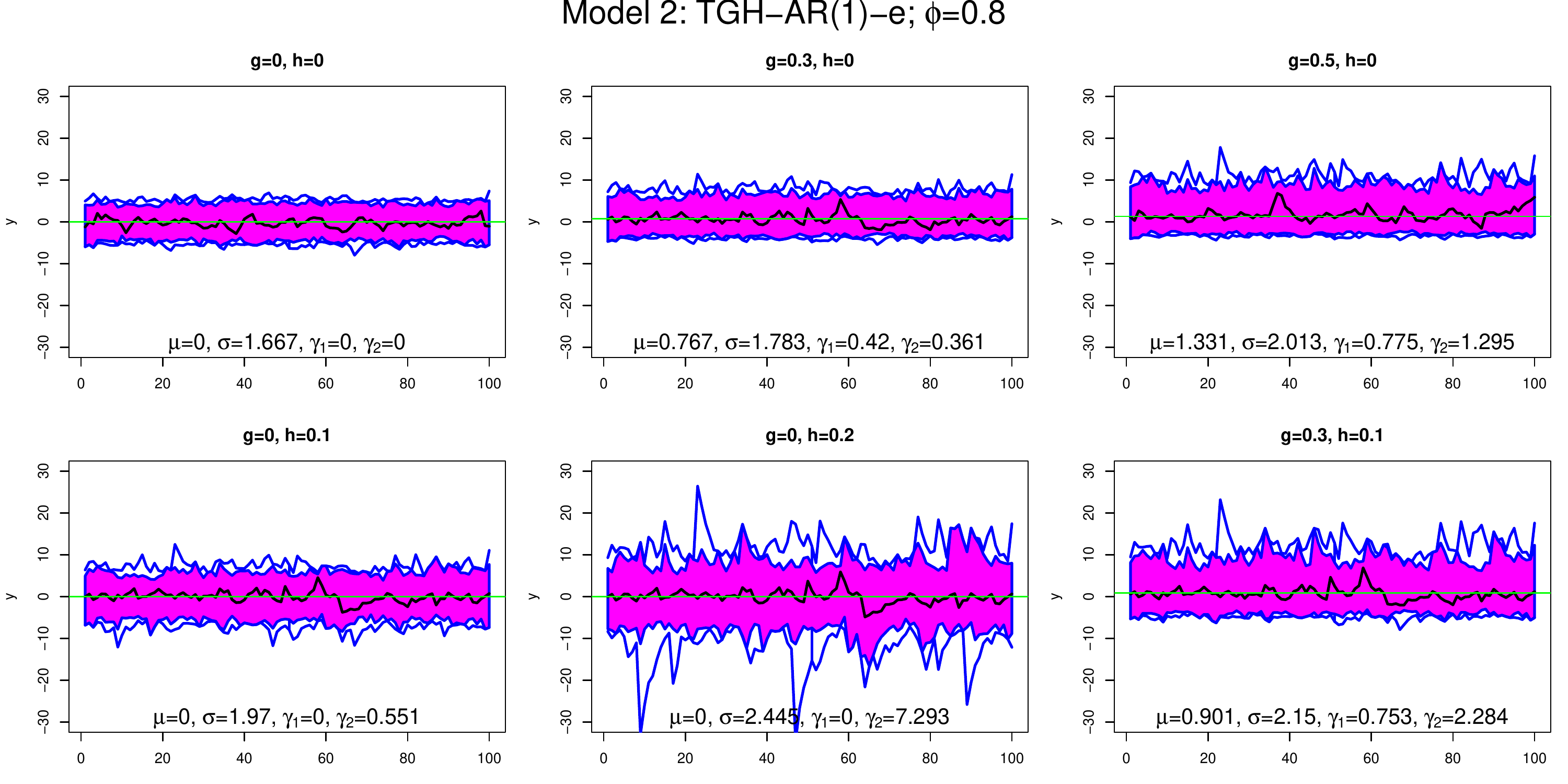}\vspace{-3mm}			
	\caption{Functional boxplots of 1000 realizations of sample size $n=100$ from model 2, i.e., TGH-AR(1)-e, without covariates, with $\xi$=0, $\omega$=1,  an AR coefficient of $\phi$=0.8, and six pairs of values for $g$ and $h$, labeled with values of the mean ($\mu$, green line), standard deviation ($\sigma$), skewness ($\gamma_1$) and excess kurtosis ($\gamma_2$) of the process.}
	\label{fb2}
\end{figure}

Instead of applying the transformation to the time series itself, in our second model we transform the error term $\epsilon_t\overset{i.i.d.}{\sim}  \mathcal{N} (0, 1)$ in an AR process to a non-Gaussian error term $T_t=\tau_{g,h}(\epsilon_t)$ that follows a standard TGH distribution. Thus, model 2 (the TGH-AR($p$)-e model) is defined as:
\begin{equation} 
	Y_t=\bm{X}_t\trans\bm{\beta}+\xi+\phi_1 \tilde{Y}_{t-1}+\cdots+\phi_p \tilde{Y}_{t-p}+\omega\tau_{g,h}(\epsilon_t), \label{model2}
\end{equation} 
where $\tilde{Y}_t=Y_t-\bm{X}_t\trans\bm{\beta}-\xi$ is a process of median 0 (however, not a zero-mean process when $g \neq 0$); $\bm{X}_t$ and $\bm{\beta}$ are the covariates and corresponding coefficients respectively; $\xi$ and $\omega$ are the location and scale parameters, respectively; and $\bm{\phi}=(\phi_1,\ldots, \phi_p)\trans$ are the AR coefficients under constraints such that the AR process is weakly stationary. We express (\ref{model2}) in a moving average form: $\tilde{Y}_t=\psi(B)(\omega T_t)$, where $\psi(B)=\sum_{j=0}^{\infty}\psi_j B^j$ of the backshift operator $B$. 
Although the marginal distribution of $Y_t$ in the TGH-AR($p$)-e model cannot be written down in a closed form, the moments of $\tilde{Y}_t$ can be derived with respect to moments of the standardized error term $T_t$: 
$\mathbb{E}(\tilde{Y}_t) = \omega\sum_{j=0}^{\infty}\psi_j \mathbb{E}(T_t) = \frac{\omega}{1-\sum_{j=1}^{p}\phi_j}\mathbb{E}(T_t), \mathbb{V}(\tilde{Y}_t) = \omega^2\sum_{j=0}^{\infty}\psi_j^2\mathbb{V}(T_t)$. Furthermore, $\gamma_1(\tilde{Y}_t) = \frac{\sum_{j=0}^{\infty}\psi_j^3}{\left(\sum_{j=0}^{\infty}\psi_j^2\right)^{3/2}}\gamma_1(T_t), \gamma_2(\tilde{Y}_t) = \frac{\sum_{j=0}^{\infty}\psi_j^4}{\left(\sum_{j=0}^{\infty}\psi_j^2\right)^{2}}\gamma_2(T_t)$, as long as the summation converges. These relationships conform with the algebraic relations between the skewness and kurtosis of the process and error term in ARMA processes  derived by \citet{Da80}.
Figure~\ref{fb2} shows the functional boxplots of 1000 realizations from the TGH-AR(1)-e model with a sample size of 100, and the same parameters as in Figure~\ref{fb1}. Again, we see that skewness and tail behavior of the process can be controlled by $g$ and $h$. 

\subsection{Comparison of the Two Models} 
Comparing Figures~\ref{fb1} and \ref{fb2}, we notice that, given the same values for $g$ and $h$, the skewness and kurtosis of the TGH-AR(1)-t process are larger than those of the TGH-AR(1)-e process, while the variance is smaller. This agrees with the findings in \citet{Da80}, where the ARMA process has skewness and kurtosis less than the error process.

\begin{figure}[t!] 
	\centering
	\includegraphics[width=\textwidth]{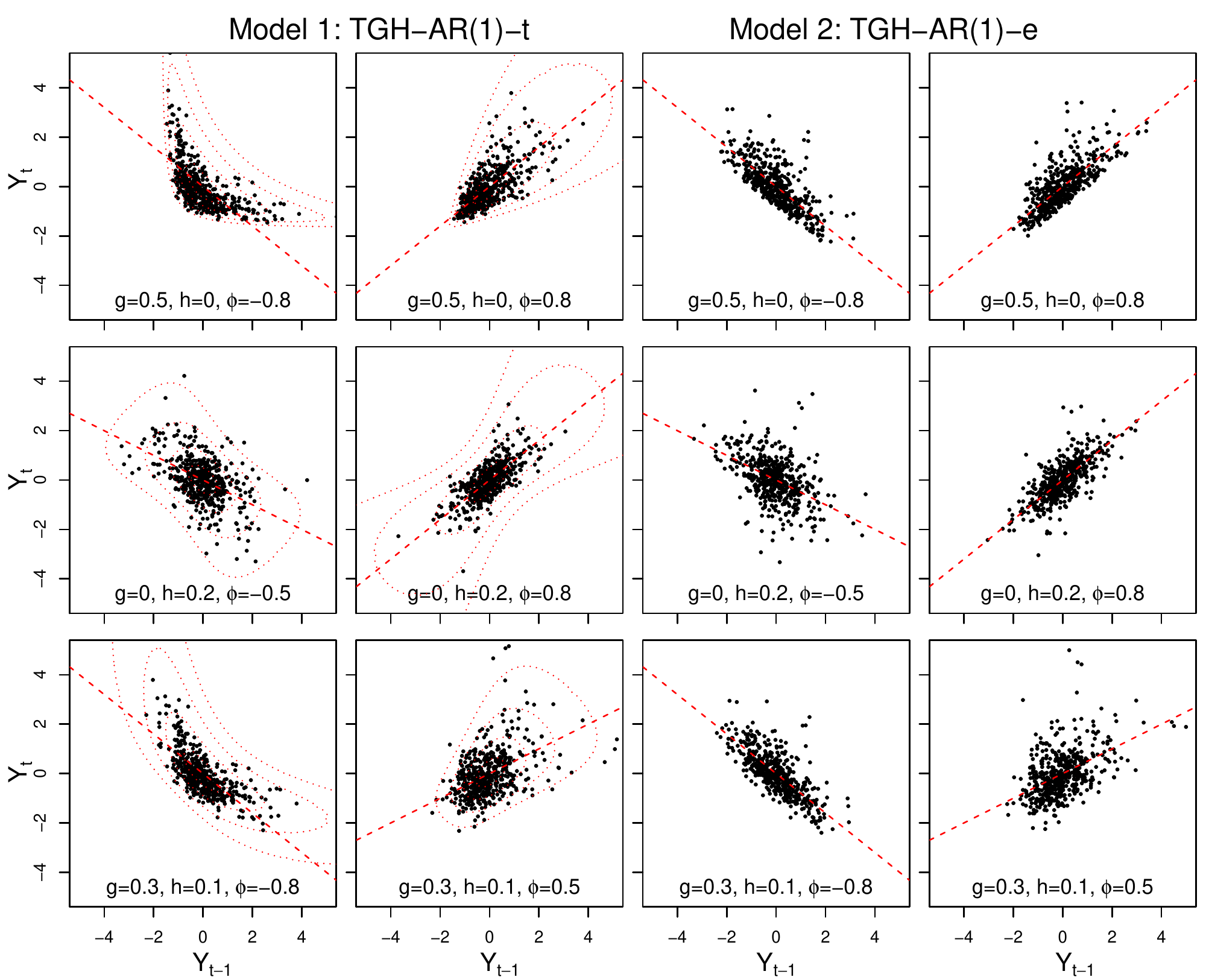}\vspace{-3mm}			
	\caption{Illustration of the non-linear relationship between $Y_{t}$ and $Y_{t-1}$ for the TGH-AR(1)-t model, overlapped with theoretical contours of the joint density (left two columns); illustration of the linear relationship between $Y_{t}$ and $Y_{t-1}$ for the TGH-AR(1)-e model with skewed and/or heavy-tailed noise (right two columns). Each panel is plotted with one realization of sample size $n=500$ from our models without covariates, with $\xi$=0, $\omega$=1 and different values for $g, h$ and $\phi$.}
	\label{rela12}
\end{figure}

We interpret the difference between the two models from the viewpoint of the data-generating mechanism. In model 1, TGH-AR-t, we assume that there is an underlying latent Gaussian process $Z_t$ and that the observed process $Y_t$ is a realization of a non-linear transformation of $Z_t$. \citet{Georg11} justified this idea using financial data. Overreactions to changes in the stock market skew the underlying symmetric process of financial news and result in skewed log-returns of the stock market; see also \citet{LucaG}. For model 2, the TGH-AR-e model, we do not assume such a latent process, but $Y_t$ itself is an AR model with non-Gaussian noise that follows the TGH distribution. The difference between the two models can be seen most obviously by plotting the relationship between $Y_t-\bm{X}_t\trans\bm{\beta}$ and $Y_{t-1}-\bm{X}_{t-1}\trans\bm{\beta}$ (Figure~\ref{rela12}). The relationship is linear for model 2 and non-linear for model 1. 
The exploratory plots shown in Figure~\ref{rela12} for a sample size of $n = 500$ and different values of $g$, $h$ and $\phi$ should be consulted as a reference for deciding whether model 1 or 2 is more suitable for the time series data to be analyzed. In practice, one should take into account both a reasonable data-generating mechanism for the process and the empirical behavior of $Y_t-\bm{X}_t\trans\bm{\beta}$ versus $Y_{t-1}-\bm{X}_{t-1}\trans\bm{\beta}$ to decide which model to use for a specific dataset.

We want to emphasize that our time series models are not simply variations of the spatial case \citep{XG17}. We utilize the unique properties of the AR structure for discrete time processes, which the random field approach does not possess. One main difference between the AR time series and the random fields is that the correlation structure of the random process is induced by the AR coefficients, whereas in the random fields, it is modeled directly. Moreover, the property of conditional independence of an AR process allows us to write down the likelihood explicitly instead of a matrix form as for the random fields. The AR structure also makes it possible for us to build the TGH-AR-e model, which is not possible in the random fields setting.

\section{Inference}\label{sec:est_pred}
\subsection{Parameter Estimation}\label{sec:est}
A drawback of the TGH transformation is that its inverse transformation does not have an explicit form (except when either $g$ or $h$ is equal to 0) and, thus, the likelihood inference is difficult. Earlier estimation procedures rely on matching sample
quantiles or sample moments with their population counterparts to avoid this difficulty. For example, \citet{Hoa85} developed an estimation procedure by matching a sequence of sample quantiles and theoretical quantiles, which we refer to as the letter-value-based method. Recently, \citet{XG15} proposed an efficient parameter estimation algorithm for the independent TGH distribution by using an approximated likelihood. Their algorithm greatly improved the parameter estimation performance compared to the moment or quantile-based methods without compromising the computational speed. 

The parameters involved in our two TGH-AR models are $\xi,\omega,g,h,\bm{\beta},\bm{\phi}$. Since the maximum likelihood estimator (MLE) is well-known to possess many good asymptotic properties, we prefer the likelihood-based estimator over the moment or quantile-based estimators. The vector of the parameters related to the TGH transformation is denoted by $\bm{\theta}_1=(\xi,\omega,g,h,\bm{\beta}\trans)\trans$,  the autoregressive parameters by $\bm{\phi}$, and set $\bm{\theta}=(\bm{\theta}_1\trans,\bm{\phi}\trans)\trans$. 

For model 1, the log-likelihood function given $n$ observations $(y_1,\bm{x}_1),\ldots,(y_n,\bm{x}_n)$ can be written as:
\be
\label{mle1}
\begin{split}
	l_n(\btheta)
	=&f_{\bm{\phi}}(z_{1,\btheta_1})+ f_{\bm{\phi}}(z_{2,\btheta_1}|z_{1,\btheta_1})+\cdots+ f_{\bm{\phi}}(z_{n,\btheta_1}|z_{n-1,\btheta_1},\ldots,z_{n-p,\btheta_1})-n\log \omega\\ &-\frac{h}{2}\sum_{t=1}^n z_{t,{\btheta_1}}^2-\sum_{t=1}^n\log\left[\exp(g z_{t,\btheta_1})+\frac{h}{g}\{\exp(g z_{t,\btheta_1})-1\} z_{t,\btheta_1}\right],
\end{split}
\ee
where $z_{t,\btheta_1}=\tau_{g,h}^{-1}\left\{\frac{y_t-\xi-\bm{x}_t\trans\bm{\beta}}{\omega}\right\}, f_{\bm{\phi}}(\cdot|\cdot)$ is the conditional log-likelihood for the underlying Gaussian AR$(p)$ process $Z_t$, which is also Gaussian.

Since an explicit form of the likelihood of $Y_t$ is intractable for model 2, we consider the conditional likelihood given the first $k\geq p$ observations. The conditional log-likelihood of model 2 can be written as: 
\be
\label{mle2}
\begin{split}
	l_n(\btheta|\,y_1,\ldots,y_k)
	=&-\frac{h+1}{2}\sum_{t=k+1}^n \epsilon^2_{t,\btheta}-\sum_{t=k+1}^n\log\left[\exp(g \epsilon_{t,\btheta})+\frac{h}{g}\{\exp(g \epsilon_{t,\btheta})-1\} \epsilon_{t,\btheta}\right]\\
	&-(n-k)\log \omega-\frac{n-k}{2}\log(2\pi),
\end{split}
\ee
where $\epsilon_{t,\btheta}=\tau_{g,h}^{-1}\left\{\frac{\tilde{y}_t-\phi_1 \tilde{y}_{t-1}-\cdots-\phi_p \tilde{y}_{t-p}}{\omega}\right\}$ and  $\tilde{y}_t=y_t-\xi-\bm{x}_t\trans\bm{\beta}$.

Although we can find the MLE for $\bm{\theta}$ in our two models by maximizing (\ref{mle1}) or (\ref{mle2}) with respect to $\bm{\theta}$ in principle, it is computationally expensive to find the inverse $\tau_{g,h}^{-1}(\cdot)$ numerically for each data point and iteration of the optimization since $\tau_{g,h}^{-1}(\cdot)$ does not have a closed form. To bypass this computational challenge, we borrow the idea of approximated likelihood from \citet{XG15} for the univariate TGH distribution. \citet{XG17} extended this idea to the random field case and proposed an algorithm for iteratively estimating both the parameters related to the TGH transformation and the parameters in the model of the covariance function. In essence, they linearized the inverse transformation function piecewisely and maximized the approximated likelihood function with the piecewisely linearized inverse function instead of the exact one.

Equipped with this linearization procedure, now we numerically maximize the approximated log-likelihood by using the approximated inverse transformation function $\tilde{\tau}_{g,h}^{-1}$ instead of $\tau_{g,h}^{-1}$ in either function~(\ref{mle1}) or (\ref{mle2}). Thus, we obtain the maximum approximated likelihood estimator (MALE) of the parameters in our models much faster. 
We also found that iteratively optimizing $\btheta_1$ and $\bm{\phi}$ is better than optimizing $\bm{\theta}$ directly; however, it is about 20 times slower on a Lenovo computer with 16  2.50GHZ Intel Xeon(R) CPUs. Hence, we optimize all of the parameters for estimation directly in our algorithm.

\subsection{Order Selection}\label{sec:ord}
The order selection for an AR model is usually performed via the Akaike information criterion (AIC) or the Bayesian information criterion (BIC). In a simulation study (all simulations are done on the same computer as mentioned above), we found that the algorithm based on BIC performs much better than the one based on AIC for our models. Therefore, we use the order selection procedure based on BIC with our estimation algorithm.
For model 1, BIC=$-2 l_n(\btheta)+(p+4)\log n$; for model 2, BIC=$-2 l_n(\btheta|\,y_1,\ldots,y_k)+(p+4)\log(n-k)$ and again we use the approximated likelihood instead of the exact one.


We evaluate the empirical correct order selection rate by BIC with the approximated likelihood for our two models through a simulation study. We generate data from both models with $\xi=0$, $\omega=1$, $g=0.3, h=0.1$ without covariates, with sample sizes $n=100$ and $n=500$ and a true AR order of $p=0,1,2$. For $p=1$ and $p=2$, we use multiple values for the AR coefficient(s), which exhibit different behaviors of the ACF and different intensities for the spectral density. We get correct order selection rate from 1000 simulations. 
Table~\ref{bic} summarizes the correct order selection rate by BIC for our two models with different sample sizes $n$ and orders $p$ for different values of $\phi$($\bm{\phi}$). 

\begin{table}[ht]
	\centering
	\caption{Correct order selection rate for both the TGH-AR($p$)-t and the TGH-AR($p$)-e models  without covariates and with $\xi=0$, $\omega=1$, $g=0.3, h=0.1$ and a true AR order of 0, 1 and 2 for $n=100$ and $n=500$.}
	\label{bic}
	\begin{tabular}{c|c|c|c|c|c}
		\hline
		&&\multicolumn{2}{c|}{TGH-AR($p$)-t}&\multicolumn{2}{|c}{TGH-AR($p$)-e}\\
		\cline{3-6}
		&& $n=100$ & $n=500$ & $n=100$ & $n=500$ \\
		\hline\hline
		$p$=0 && 0.959 & 0.988 & 0.938 & 0.990 \\ \hline 
		\multirow{4}{*}{$p$=1} & $\phi=-0.5$ & 0.951 & 0.987 & 0.934 & 0.987\\
		 &$\phi=0.5$& 0.950 & 0.988 & 0.932 & 0.981 \\ 
		 &$\phi=-0.8$& 0.952 & 0.982 & 0.940 & 0.985 \\ 
		 &$\phi=0.8$& 0.961 & 0.987 & 0.942 & 0.992 \\ \hline 
		\multirow{4}{*}{$p$=2} & $\bm{\phi}=(-0.5,-0.3)\trans$ & 0.825 & 0.991 & 0.858 & 0.987\\
		&$\bm{\phi}=(0.2,0.4)\trans$& 0.532 & 0.986 & 0.592 & 0.986 \\ 
		&$\bm{\phi}=(0.8,-.25)\trans$& 0.706 & 0.987 & 0.777 & 0.983 \\ 
		&$\bm{\phi}=(1.5,-0.75)\trans$& 0.948 & 0.982 & 0.937 & 0.983 \\ \hline 
		\hline		
	\end{tabular}
\end{table}

The correct order selection rate improved and reached over 98\% by increasing the sample size from 100 to 500 in all cases. Order selection performance is similar with data generated by all the six pairs of values for $g$ and $h$ as used in the functional boxplots. We conclude that the overall correct order selection rate is satisfactory for our order selection procedure based on BIC with the approximated likelihood. 

\subsection{Forecasting}\label{sec:pred}
In this section, we derive one-step-ahead point and probabilistic forecasts for our two models. 

For model 1, $Y_t=\bm{X}_t\trans\bm{\beta}+\xi+\omega T_t$, given $p$ observations $y_{t},\ldots,y_{t-p+1}$, we derive the conditional distribution: 
$$T_{t+1} |y_{t},\ldots,y_{t-p+1} \sim \tau_{g,h}(\tilde{\mu}+\tilde{\sigma}Z),\, Z\sim \mathcal{N}(0,1),$$
where $\tilde{\mu}$ and $\tilde{\sigma}^2$ are the conditional mean and variance, respectively, for $Z_{t+1}|Z_{t},\ldots,Z_{t-p+1}$ of the underlying Gaussian AR($p$) process $Z_t$. Here, $\tilde{\mu}$ and $\tilde{\sigma}^2$ are determined by $\bm{\phi}$ and can be found efficiently by the Durbin-Levinson algorithm. With this conditional distribution, the point predictors for minimizing the absolute prediction error (conditional median) and the squared prediction error (conditional mean) are: $
\hat{Y}_{t+1}=\xi+\bm{X}_{t+1}\trans\bm{\beta}+\omega\tau_{g,h}(\tilde{\mu})$
and $ \hat{Y}_{t+1}=\xi+\bm{X}_{t+1}\trans\bm{\beta}+\frac{\omega}{g\sqrt{1-h\tilde{\sigma}^2}}\exp\left\{\frac{h\tilde{\mu}^2}{2(1-h\tilde{\sigma}^2)}\right\}\left[\exp\left\{\frac{g^2\tilde{\sigma}^2+2g\tilde{\mu}}{2(1-h\tilde{\sigma}^2)}\right\}-1\right]$,
respectively. 

For model 2, the conditional distribution is: 
$$Y_{t+1} |y_{t},\ldots,y_{t-p+1} \sim \bm{X}_{t+1}\trans\bm{\beta}+\xi+\phi_1 \tilde{y}_{t}+\cdots+\phi_p \tilde{y}_{t-p+1}+\omega\tau_{g,h}(Z),\, Z\sim \mathcal{N}(0,1).$$ The point predictors for minimizing the absolute loss (conditional median) and squared loss (conditional mean) are 
$\hat{Y}_{t+1}=\xi+\bm{X}_{t+1}\trans\bm{\beta}+\phi_1 \tilde{y}_{t}+\cdots+\phi_p \tilde{y}_{t-p+1}$ and $ \hat{Y}_{t+1}=\xi+\bm{X}_{t+1}\trans\bm{\beta}+\phi_1 \tilde{y}_{t}+\cdots+\phi_p \tilde{y}_{t-p+1}+\frac{\omega}{g\sqrt{1-h}}\left[\exp\left\{\frac{g^2}{2(1-h)}\right\}-1\right]$, respectively. We note that the conditional median has the same form as that of an AR model with Gaussian error. In practice, we need to estimate the parameters first to make forecasts for future values of the time series based on those estimated parameters. Thus, the difference between the point predictions based on the conditional medians of our model and the Gaussian AR model comes from the difference in their respective estimations of $\bm{\beta}, \xi$ and $\bm{\phi}$. 

Prediction confidence intervals (CI) can be found easily from the conditional distributions of our two models. There are different ways to form the two-sided CI for a given distribution. One popular choice for the $(1-\alpha)$100\% CI is to exclude $\alpha/2$ from both tails and then use the central $1-\alpha$ probability interval of the distribution, which we refer to as the symmetric weight CI. Another choice is to find the shortest $1-\alpha$ probability interval, which we refer to as the minimum-length CI. The minimum-length CI coincides with the symmetric weight CI for symmetric distributions. For model 1, the lower and upper bounds of the 95\% prediction CI can be found by transforming the lower and upper bounds of a 95\% probability interval of the underlying normal distribution of mean $\tilde{\mu}$ and variance $\tilde{\sigma}^2$. The symmetric weight prediction interval is $[\bm{X}_{t+1}\trans\bm{\beta}+\xi+\omega \tau_{g,h}(\tilde{\mu}-z_{1-\alpha/2}\tilde{\sigma}),\bm{X}_{t+1}\trans\bm{\beta}+\xi+\omega \tau_{g,h}(\tilde{\mu}+z_{1-\alpha/2}\tilde{\sigma})]$, and the minimum-length prediction interval is $[\bm{X}_{t+1}\trans\bm{\beta}+\xi+\omega \tau_{g,h}(\tilde{\mu}-z_{1-\gamma^{\text{opt}}}\tilde{\sigma}),\bm{X}_{t+1}\trans\bm{\beta}+\xi+\omega \tau_{g,h}(\tilde{\mu}+z_{1-\alpha+\gamma^{\text{opt}}}\tilde{\sigma})]$, where $\gamma^{\text{opt}}$ needs to be optimized numerically for different values of $g$ and $h$ for given $\tilde{\mu}$ and $\tilde{\sigma}$. 
For model 2, the symmetric weight prediction interval is $[\hat{Y}_{t+1}+\omega \tau_{g,h}(-z_{1-\alpha/2}),\hat{Y}_{t+1}+\omega \tau_{g,h}(z_{1-\alpha/2})]$ and the minimum-length prediction interval is $[\hat{Y}_{t+1}+\omega \tau_{g,h}(-z_{1-\gamma^{\text{opt}}}),\hat{Y}_{t+1}+\omega \tau_{g,h}(z_{1-\alpha+\gamma^{\text{opt}}})]$, where $\hat{Y}_{t+1}$ is the conditional median of $Y_{t+1} |y_{t},\ldots,y_{t-p+1}$. 

The prediction intervals given above are derived with respect to the true parameters of the two models. However, in practice, those parameters need to be estimated and the prediction interval with the estimated parameters should be inflated taking the uncertainty in parameter estimation into account. Research on mean squared prediction error (MSPE) with estimated parameters can be referred to a vast literature in the context of linear models, time series models and spatial models \citep{MSPESpZC92}. In particular, for a Gaussian AR($p$) model, the MSPE of a one-step ahead forecast with estimated autoregressive coefficients is inflated by a factor of $1 + p/n$, where $n$ is the sample size \citep{MSPEAR76}. Many authors \citep[e.g.][]{MSPELMToy82} have concluded that the bias of the estimated MSPE is asymptotically negligible and we see later in the simulation study that the empirical coverage of the prediction CIs by ignoring the uncertainty in parameter estimation is close to the nominal level.

\section{Simulation Study}\label{sec:sim}
We perform a Monte Carlo simulation study to assess performance of our models from different aspects. We consider six pair of values for $g$ and $h$ (same values as in the functional boxplots) and sample sizes $n=100, 250, 500$. For each of our TGH-AR(1)-t and TGH-AR(1)-e model, each sample size $n$ and each pair of values for $g$ and $h$, we first generate one realization from the model with $\xi=-3$, $\omega=1.5, \bm{\beta}=(3,-2)\trans, \bm{X}_t=\left\{\cos(2\pi t/24),\sin(2\pi t/24)\right\}\trans$ and $\phi=0.8$. Next, we carry out parameter estimation by three methods: the MALE for both of our models without order selection as well as the MLE for a Gaussian AR model without order selection. Finally, we use again the three methods based on the TGH-AR-t, TGH-AR-e and Gaussian AR models, to produce point and probabilistic forecasts with the estimated parameters. For each scenario, we run the above procedure 1000 times.

In Section~\ref{sec:sim1}, we compare the MALE for the TGH-AR(1)-t model with two other estimators for the independent TGH distribution when data are generated from the TGH-AR(1)-t model. In Section~\ref{sec:sim2} we evaluate the finite sample behavior of the MALE for both models and in Section~\ref{sec:sim3} we compare the forecasting performances of our models with the performance of a Gaussian AR model. 

\subsection{Comparison of Estimation Methods}\label{sec:sim1}
Using the TGH-AR-t model, we first check the improvement of estimation performance by using the MALE to simultaneously estimate $\bm{\theta}_1$ and $\bm{\phi}$ rather than sequentially estimating $\bm{\theta}_1$ from $Y_t$ treating them as independent realizations and then estimate $\bm{\phi}$ from the transformed process 
$\tilde{Z_t}=\tau^{-1}_{\hat{g},\hat{h}}(\frac{Y_t -\hat{\xi}-\bm{X}_{t}\trans\hat{\bm{\beta}}}{\hat{\omega}})$. The latter approach is often adopted by practitioners to first find the optimal transformation and then estimate the temporal structure from the `Gaussianized' process. 
For this comparison, we use two additional methods for estimation in the simulation procedure as described before: the letter-value-based method \citep{Hoa85} and the MALE for independent TGH distributions \citep{XG15} to estimate $\bm{\theta}_1$ first and then estimate $\phi$ by the transformed process $\tilde{Z_t}$.

\begin{figure}[t!] 
	\centering
	\begin{tabular}{c}
		\includegraphics[width=\textwidth]{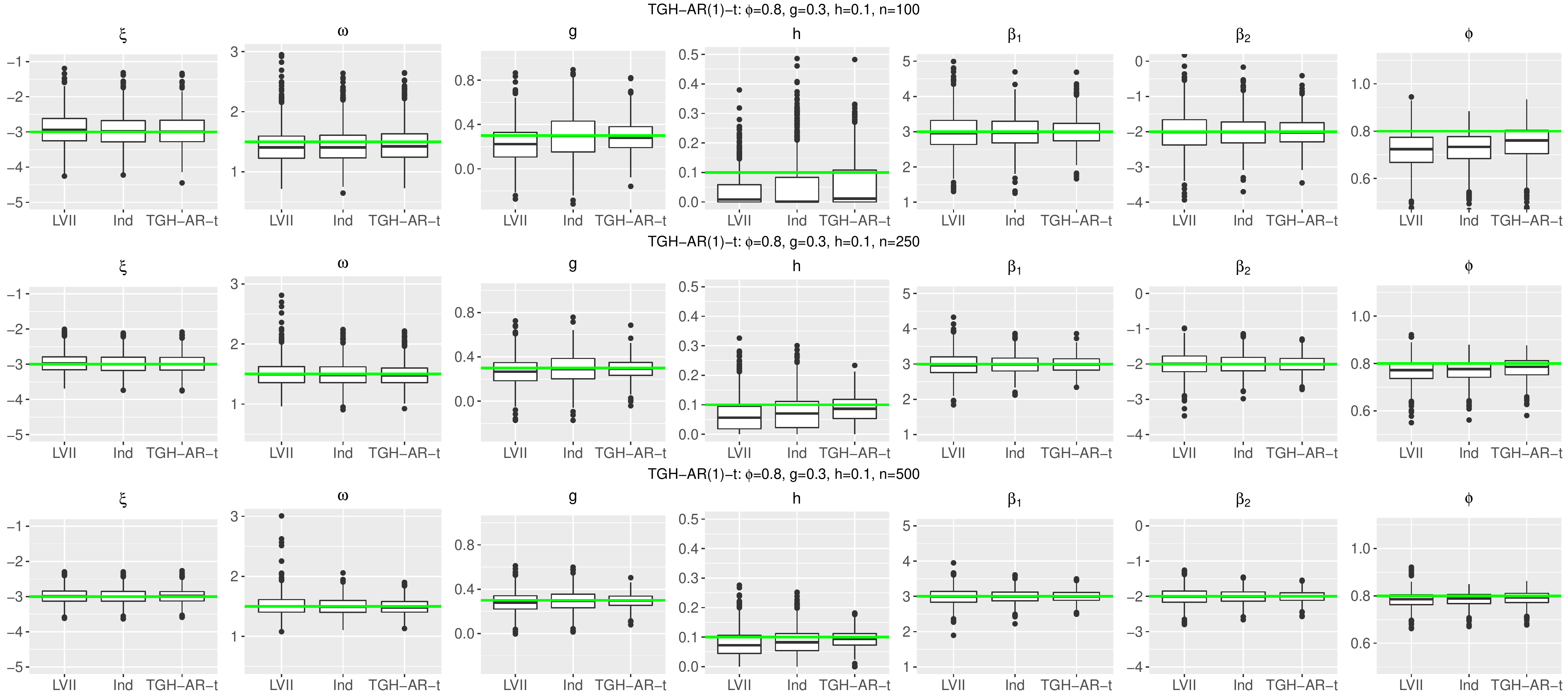}	
	\end{tabular}		
	\caption{Comparison of three estimators: letter-value-based method (LVII), MALE based on independent TGH distribution (Ind) and MALE for the TGH-AR-t model (TGH-AR-t), with boxplots of each estimator for each parameter from 1000 replications with data generated from the TGH-AR($1$)-t with true parameters indicated by green line.}
	\label{compare}
\end{figure}

Figure~\ref{compare} presents boxplots of the parameters estimated by the three different estimators for 1000 replications with $g=0.3, h=0.1$ for 3 sample sizes. We see that the bias and variance of the estimators improve dramatically as the sample size grows from 100 to 500 for all three methods. Our joint estimation procedure outperforms the other two sequential methods, especially for $h$ and $\phi$, which indicate that estimating the optimal transformation by ignoring the dependency structure of the underlying process deteriorate the estimation. Hence, the comparison result justifies the need to develop such a tailored estimation procedure for the TGH distribution in the time series context. 

\subsection{Estimation Performance of MALE}\label{sec:sim2}

\begin{figure}[t!] 
	\centering
	\includegraphics[width=\textwidth]{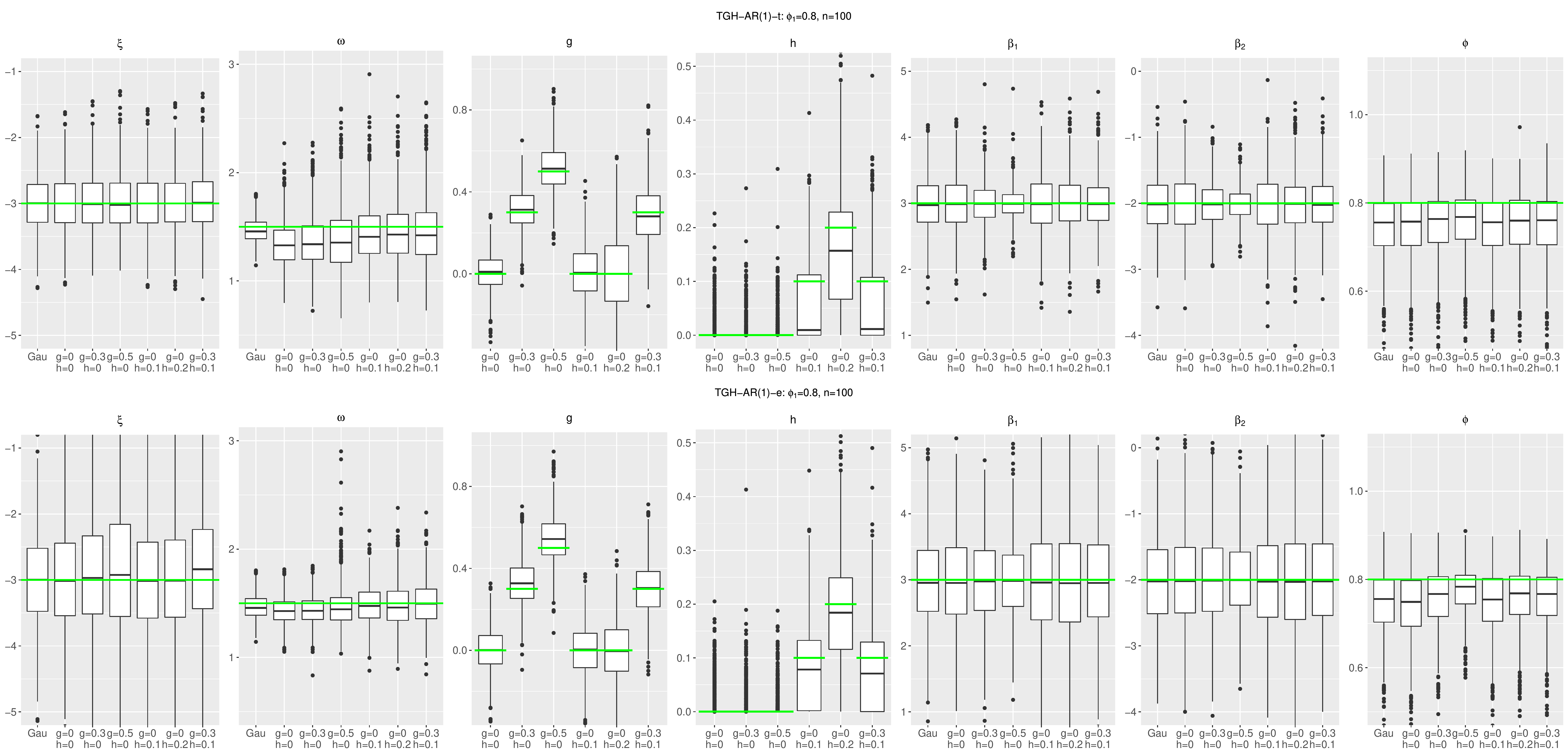}	
	\caption{Boxplots of the MALE in 1000 simulations for the two TGH-AR(1) models with $n=100$ for six pairs of different values of $g$ and $h$ for each parameter with true value indicated by green line. For $g=h=0$, boxplots of the MLE based on the Gaussian AR model (indicated by `Gau') for $\xi,\omega,\beta_1,\beta_2,\phi$ are also included.}
	\label{est}
\end{figure}

Next, we check the finite sample behavior of the MALE for both models.
Figure~\ref{est} shows the estimation results from our two TGH-AR(1) models with $\phi=0.8$, $n=100$ for six pairs of different values of $g$ and $h$. For $g=h=0$, which corresponds to a Gaussian AR process, we also include boxplots of the MLE based on the Gaussian AR model for $\xi,\omega,\beta_1,\beta_2$ and $\phi$. Visunimations of boxplots in a same manner as sample size grows ($n=100, 250, 500$) for both models can be found in the supplementary material (Visuanimations~S2 and S3). Results of estimation performance for the two TGH-AR(2) models with $\phi_1=0.8, \phi_1=-0.25$ are also included in the supplementary material (Visuanimations~S4 and S5).

From Figure~\ref{est}, we can see that with sample size $n=100$, the MALE for $\omega, h$ and $\phi$ is a bit biased. Nevertheless, as shown by Visuanimations~S2 and S3 in the supplementary material, the estimation improves greatly as the sample size increases to $n=250$. In addition, under $g=h=0$ when the true distribution is Gaussian, with two more parameters, $g$ and $h$, to estimate, the estimators based on the approximated TGH-AR likelihood do not deteriorate much from the Gaussian MLE. The visuanimations also give us an empirical guideline about how the sample size affects the estimation performance. To get an overall satisfying estimation performance by the MALE, we suggest a sample size no less than $n=250$.


%

\subsection{Forecasting Performance}\label{sec:sim3}
Finally, we evaluate the forecasting performance for our two models with parameters estimated by the corresponding MALE without order selection. 
Table~\ref{pred_t} summarizes the performance of the point forecasts from each of the three methods that based on the TGH-AR-t, TGH-AR-e and Gaussian AR models, and for data generated from each of our two TGH-AR(1) models with $\xi=-3$, $\omega=1.5, \bm{\beta}=(3,-2)\trans, \bm{X}_t=\left\{\cos(2\pi t/24),\sin(2\pi t/24)\right\}\trans$ and $\phi=0.8$ for $g=0.3, h=0.1$ and $n=500$. The mean absolute errors (MAE) is calculated from the conditional median and the root mean square errors (RMSE), from the conditional mean as point predictor. It also shows the empirical coverage and average length of the 95\% minimum-length CI from each method.

\begin{table}[ht]
	\caption{Summary of point forecast performance of the three methods based on the TGH-AR-t, TGH-AR-e and Gaussian AR models, and for data generated from each of our two TGH-AR(1) models with $\xi=-3$, $\omega=1.5, \bm{\beta}=(3,-2)\trans, \bm{X}_t=\left\{\cos(2\pi t/24),\sin(2\pi t/24)\right\}\trans$, $\phi=0.8$ for $g=0.3, h=0.1$, $n=500$.}
	\centering
	\begin{tabular}{c||c|c|c||c|c|c}
		\hline
		Data generated from &\multicolumn{3}{c|}{Model 1: TGH-AR(1)-t}&\multicolumn{3}{|c}{Model 2: TGH-AR(1)-e}\\
		\cline{1-7}
		Modeled by & TGH-AR-t & TGH-AR-e & Gau-AR & TGH-AR-t & TGH-AR-e & Gau-AR\\ 
		\hline\hline
		MAE & \textbf{0.860} & 0.869 & 0.870 & 1.374 & \textbf{1.365} & 1.379\\ \hline 
		RMSE & \textbf{1.124} & 1.142 & 1.139 & 1.798 & \textbf{1.795} & 1.801 \\ \hline 
		95\% CI coverage & \textbf{95.6\%} & 95.2\% & 94.6\% & 95.0\% & \textbf{95.6\%} & 95.6\% \\ \hline
		95\% CI width & \textbf{4.30} & 4.60 & 4.68 & 7.40 & \textbf{7.24} &7.55
	\end{tabular}
	\label{pred_t}
\end{table}	

\begin{figure}[p] 
	\centering
	\includegraphics[width=0.9\textwidth]{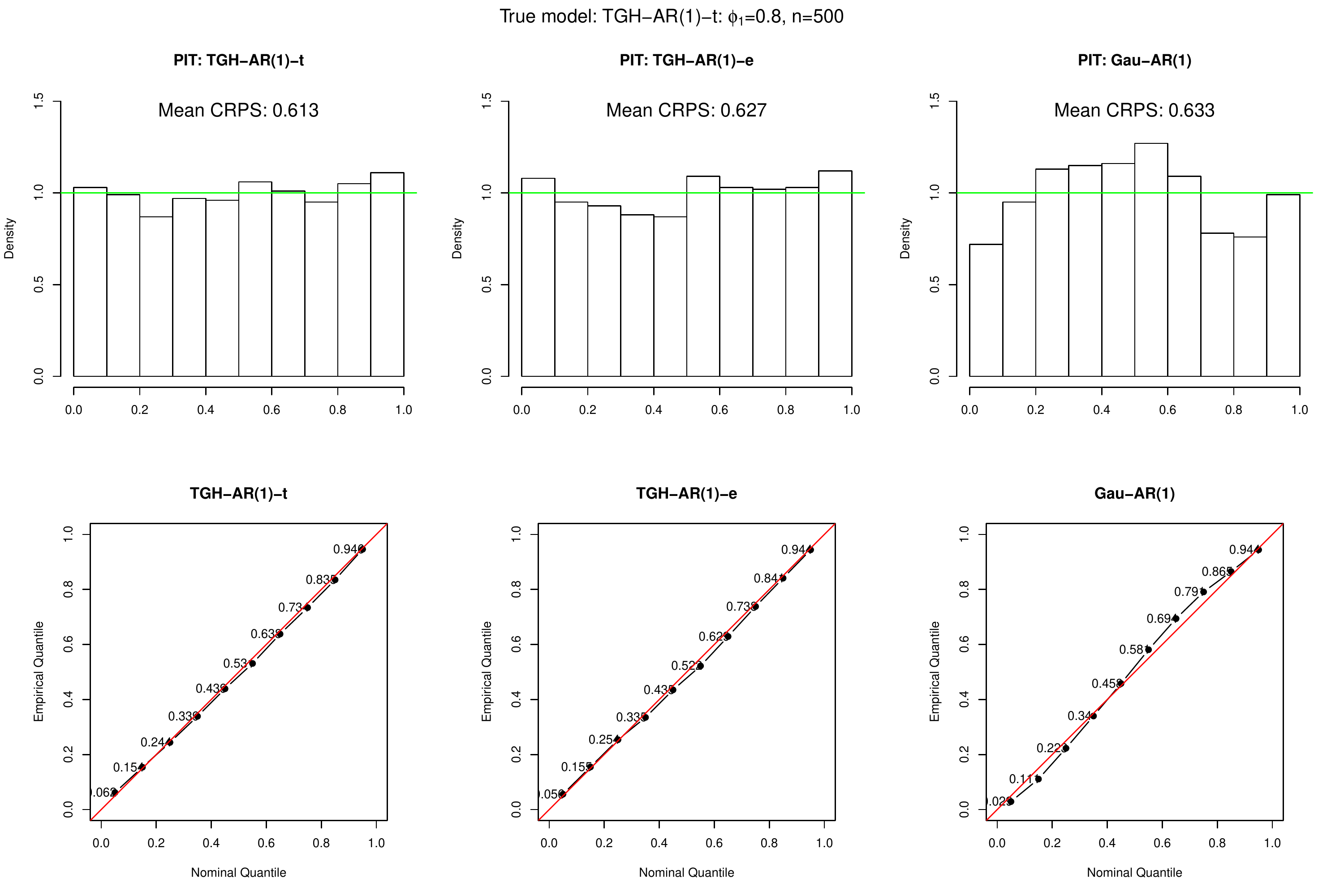}
	\vspace{-5mm}
	\caption{Comparison of probabilistic forecast performances via PIT and reliability plots when data are generated from TGH-AR(1)-t model with $\xi=-3$, $\omega=1.5, \bm{\beta}=(3,-2)\trans, \bm{X}_t=\left\{\cos(2\pi t/24),\sin(2\pi t/24)\right\}\trans$, $\phi=0.8$ for $g=0.3, h=0.1$, $n=500$ and fitted for the three methods: TGH-AR(1)-t, TGH-AR(1)-e and Gau-AR(1). Mean CRPS value is also labeled.}
	\label{pred1}
	\includegraphics[width=0.9\textwidth]{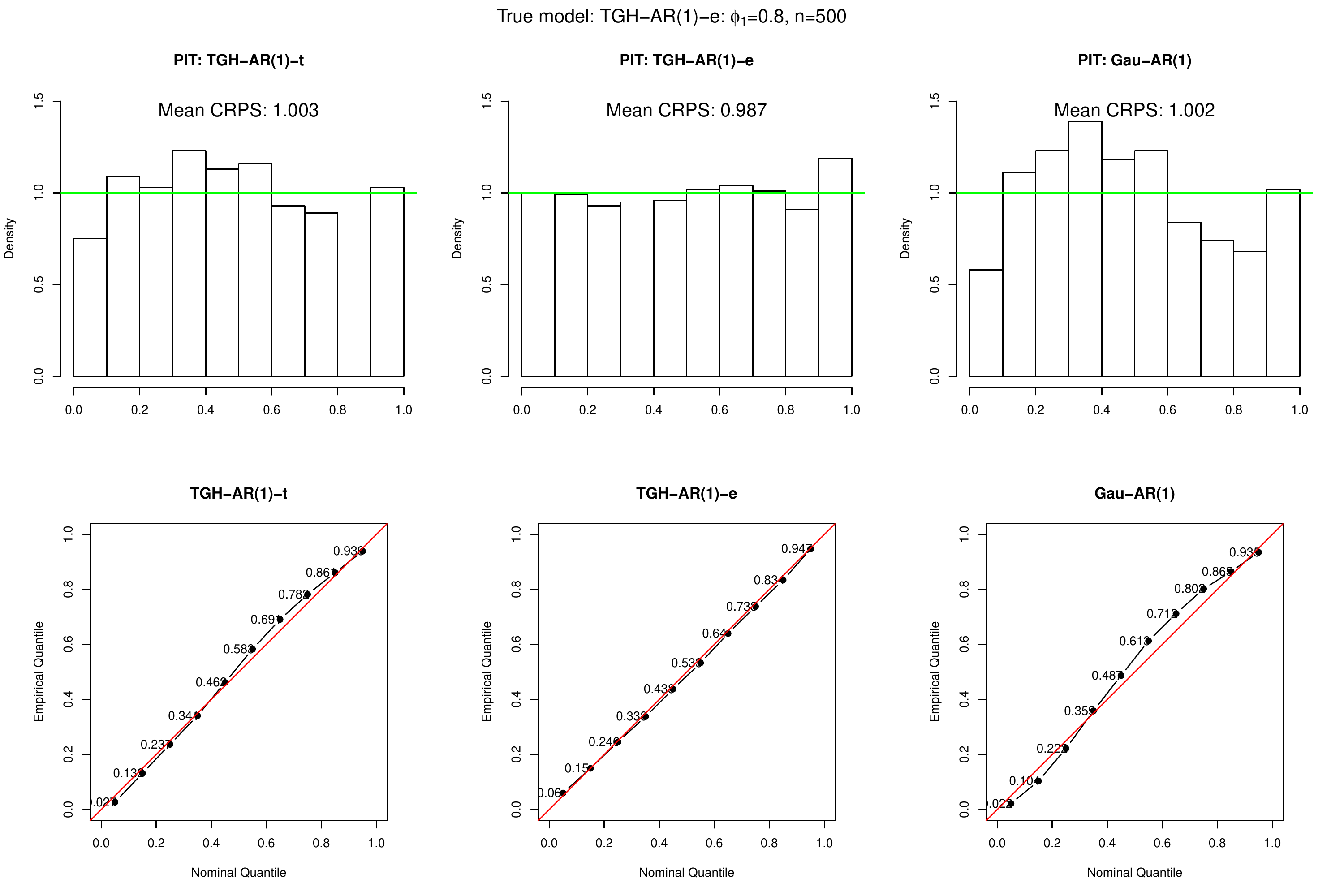}
	\vspace{-5mm}
	\caption{Comparison of probabilistic forecast performances via PIT and reliability plots when data are generated from TGH-AR(1)-e model with $\xi=-3$, $\omega=1.5, \bm{\beta}=(3,-2)\trans, \bm{X}_t=\left\{\cos(2\pi t/24),\sin(2\pi t/24)\right\}\trans$, $\phi=0.8$ for $g=0.3, h=0.1$, $n=500$ and fitted for the three methods: TGH-AR(1)-t, TGH-AR(1)-e and Gau-AR(1). Mean CRPS value is also labeled.}	
	\label{pred2}
\end{figure}

The probabilistic forecasts of the three methods can be evaluated through a histogram of the probability integral transform (PIT) values, which is to apply the conditional cumulative distribution function to the true value of the process at the time point for forecasting. If the conditional distribution assumed by a certain model conforms with the true conditional distribution, the histogram should be flat (uniform). Another metric for evaluating probabilistic forecast is the continuous ranked probability score (CRPS); more on probabilistic forecasts and CRPS for a Gaussian distribution can be found in \citet{GnKa} and \citet{GM06}. \citet{XG17} derived the CRPS for a TGH distribution. The lower the CRPS is, the better the probabilistic forecast is. A plot of the empirical versus nominal quantile, i.e., a reliability plot, can be used to check the quality of the quantile prediction. 
Figures~\ref{pred1} and \ref{pred2} show histograms of the PIT values (labeled with the mean CRPS value) and reliability plots for all three forecasting methods applied to the data generated from each of our TGH-AR(1) model. Figures S1 and S2 in the supplementary material show probabilistic forecast results for the two TGH-AR(2) models with $\phi_1=0.8, \phi_1=-0.25$.

Even though in Table~\ref{pred_t}, the MAEs and RMSEs do not differ much between the models, superiority of a model can be seen evidently by comparison of the PIT histograms. Not surprisingly, the best forecast is achieved by using the same model that the data are generated from: the forecast has the smallest MAE and RMSE, a flatter PIT histogram, and a smaller mean CRPS; the reliability plot is close to a $45^{\circ}$ straight line; the empirical coverage of the 95\% CI is close to the nominal level while the width is shorter than that of the Gaussian AR predictor. 
Note that, although we used the prediction CIs derived in Section~\ref{sec:pred} by treating the estimated parameters as the true parameters, Table~\ref{pred_t} shows that the empirical coverage of the 95\% CIs is close to the nominal level. We also notice that the TGH-AR-t and TGH-AR-e models produce quite different forecasts via their distinctive histograms of the PIT values, which means that the two models are not interchangeable.
\section{Application to Wind Speed Data}\label{sec:app}
The analysis of wind data is a crucial step in simulations for climate science. The accurate forecasting of wind speeds and quantifying the forecast uncertainty are also important for exploiting wind as clean energy. In this section, we illustrate the usefulness of our two non-Gaussian time series models under both wind speed simulation and wind speed forecasting scenarios. 
In Section~\ref{sec:app1} we fit daily wind speed data with the TGH-AR-t model to get parameter estimation for the purpose of fast wind speed simulation. In Section~\ref{sec:app2}, we apply the TGH-AR-e model in order to make better wind speed forecasts.
For the two datasets in the two subsections, the reason to use one specific model is based on plots of $Y_t-\bm{X}_t\trans\bm{\beta}$ versus $Y_{t-1}-\bm{X}_{t-1}\trans\bm{\beta}$ and with reference to Figure~\ref{rela12} to see whether the relationship is linear or not.

\subsection{Wind Speed Simulation}\label{sec:app1}
%
Climate models can produce multiple outputs of spatio-temporal wind speed data over the globe. Statistical models have been developed to reproduce the output from climate models for the sake of fast simulation instead of using the computationally expensive physical models. In order to fit a space-time statistical model for wind field over the entire world, a 4-step multi-resolution method has been proposed by \citet{CG16}, in which the first step is to model the time series of wind speed at each location individually; see \citet{CG18} for the general principles for analyzing big spatio-temporal data from climate models. For the 4-step multi-resolution method, our TGH-AR time series models can be used as a modification of the first step when time series data show non-Gaussian features.

\begin{figure}[b!] 
	\centering
	\includegraphics[width=\textwidth]{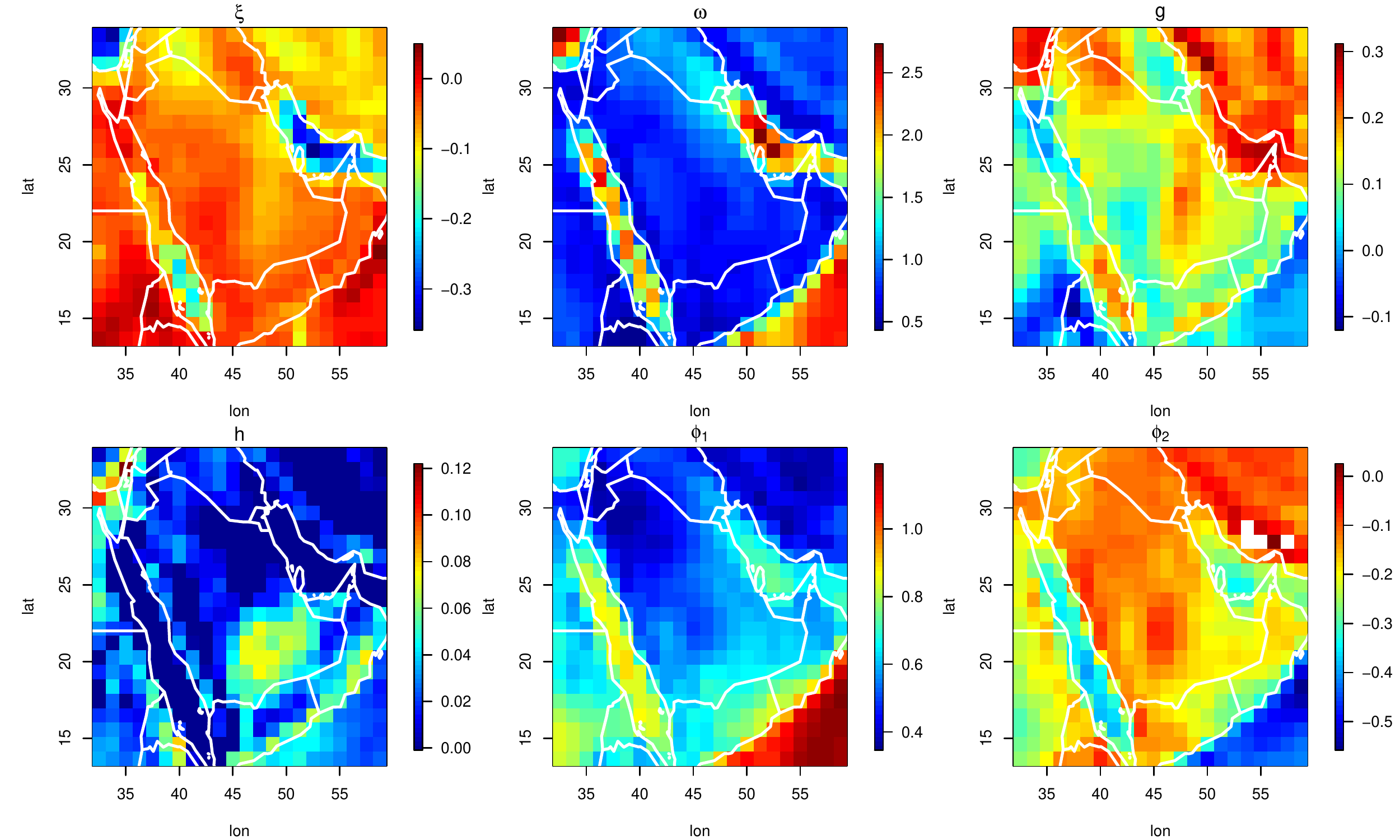}
	\caption{Maps of the estimated parameters from fitting the TGH-AR-t model with order selection to daily wind speed residuals.}
	\label{par1}
\end{figure}

To illustrate the usefulness of our TGH-AR-t model, we consider a publicly available Large Ensemble Project (LENS) dataset that consists of 30 ensembles of daily wind speed over the globe with a spatial resolution of around $1^\circ$ longitude and  $1^\circ$ latitude from the year 1920 to 2100 \citep{Kay15}. We select one ensemble from the complete dataset and use the historical 86 years (1920-2005, $n=31390$) at 22 $\times$ 22 gridded locations over Saudi Arabia (bounded roughly by $13-34^{\circ}$N and $32-59^{\circ}$E). For each day in a year for each location, we estimate the seasonality by taking the average of the wind speed across the 86 years at that location. For each location, we remove the seasonal effect from each time point and analyze the residual wind speed data, which show a clear AR pattern by the ACF and the partial autocorrelation function (PACF). We use the TGH-AR-t model because the residual wind speed time series clearly show features of skewness and heavy tails and a plot of $Y_t$ versus $Y_{t-1}$ for the residual wind speed shows a nonlinear relationship. We get parameter estimation with order selection by fitting the TGH-AR-t model at each location. With the TGH-AR-t model and the estimated parameters at each location, wind speed data can be generated rapidly without using the physical model. The estimated parameters also give insights into the pattern of the distributional properties for the wind speed residuals.


Figure~\ref{par1} shows maps of the estimated values of the 4 parameters related to the Tukey $g$-and-$h$ transformation as well as two autoregressive coefficients. Since the autoregressive order selected is 2 for the majority of the locations, maps of estimations of higher order autoregressive coefficients are omitted. We may notice from these plots that $\hat{\xi}$, $\hat{\omega}$, $\hat{\phi_1}$ and $\hat{\phi_2}$ (white area indicate the order selected is only 1) are distinctively different between land and ocean. We also observe that the $\hat{g}$ and $\hat{h}$ estimates show interesting patterns that are closely related to elevation and geographical features of a location. 
For the full 4-step multi-resolution analysis of daily wind speed over the globe, where the TGH-AR-t model is used in the first step, see \citet{JY18}.

\subsection{Hourly Wind Speed Data at a Meteorological Station}\label{sec:app2}
We consider hourly wind speed data observed at a meteorological tower in Sunnyside, Oregon, from 1 December 2013 to 31 December 2014 \citep{KH15}. 
First, we use the hourly wind speed observed in June 2014 ($n=720$) of this dataset to demonstrate the suitability of using the TGH-AR-e model. We notice that there exists a diurnal pattern in the hourly wind speed, so we include harmonics with periods of 12 and 24 hours as the covariates in the model. We find the MALE with order selection by fitting the wind speed time series from that month with the TGH-AR-e model. The estimated parameters are $\hat{\xi}=7.84,\, \hat{\omega}=1.65,\, \hat{g}=0.11,\, \hat{h}=0.06,\, \hat{\phi}_1=1.01,\, \hat{\phi}_2= -0.15$, and $\hat{\bm{\beta}}=(-0.48, 2.31, -0.14, 0.19)\trans$ with $\bm{X}_t=\left\{\cos(2\pi t/12),\sin(2\pi t/12),\cos(2\pi t/24),\sin(2\pi t/24\right\}\trans$. 
\begin{figure}[t!] 
	\includegraphics[width=\textwidth]{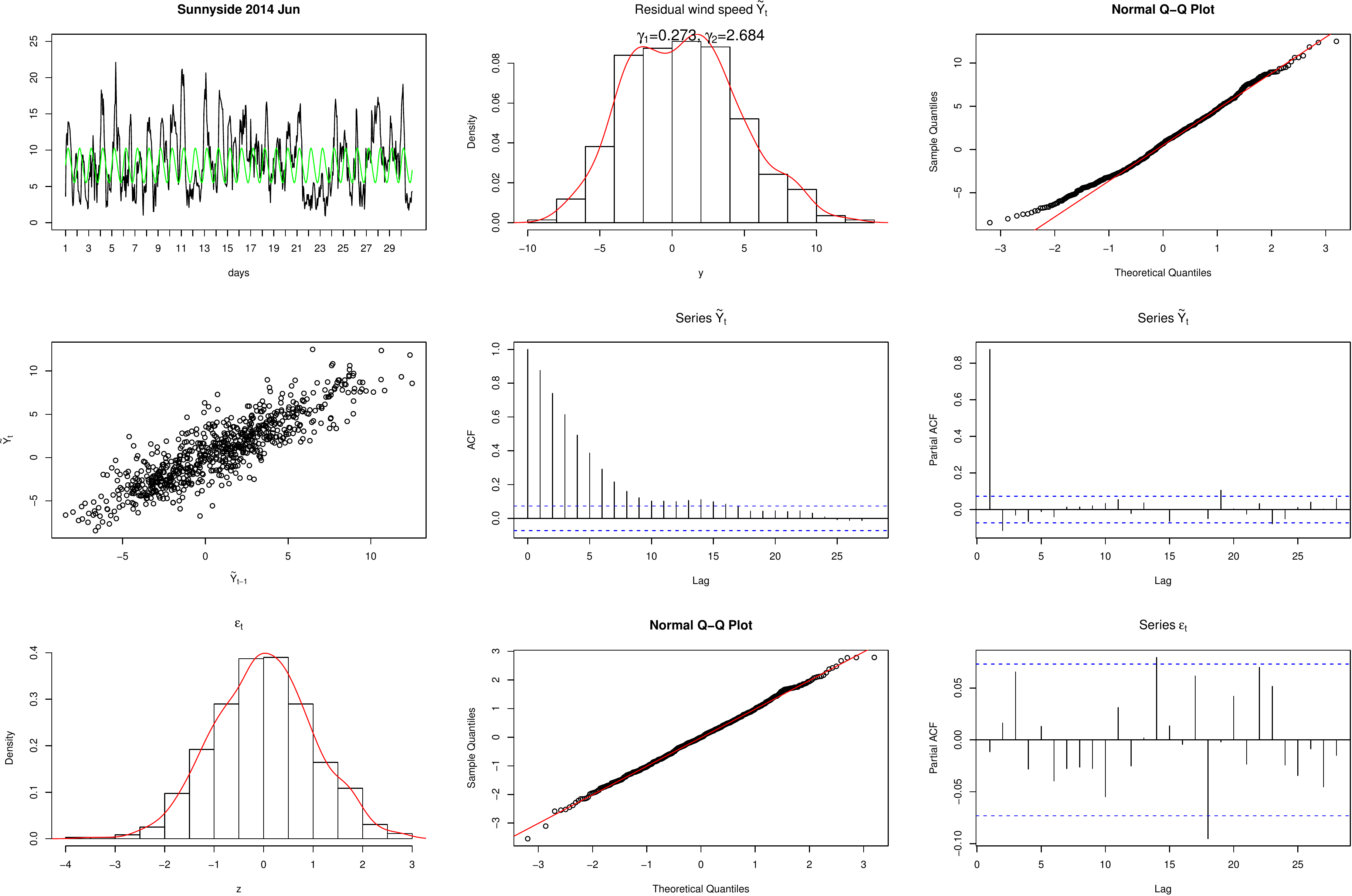}	
	\caption{Various diagnostic plots of applying the TGH-AR-e model to the hourly wind speed data of the month June 2014.}
	\label{wind1}
\end{figure}

Figure~\ref{wind1} shows the original hourly wind speed time series overlapped with a diurnal pattern (green line) estimated using the TGH-AR-e model. The histogram and normal Q-Q plot of the residual wind speeds after removing periodicity obviously deviate from Gaussianity. The skewness and kurtosis values presented with the histogram further support using our TGH-AR-e model to adapt to right-skewed and heavy-tailed data. The residual wind speed at time $t$ is plotted against time $t - 1$ in the first panel of the middle row in Figure~\ref{wind1}, showing a strong linear relationship, which is the reason why we choose to transform the error term using the TGH-AR-e model rather than the process itself. The ACF and the PACF indicate an absence of seasonality after removing the diurnal pattern with the harmonics by the TGH-AR-e model. The ACF and the PACF also validate the appropriateness of using an AR(2) model, selected by BIC, for the wind speed residuals process.
The bottom row of Figure~\ref{wind1} shows a histogram, a normal Q-Q plot and the PACF for the estimated back-transformed Gaussian error term  $\tilde{\epsilon}_{t,\btheta}=\tau_{\hat{g},\hat{h}}^{-1}\left\{\frac{\hat{\tilde{y}}_t-\hat{\phi_1} \hat{\tilde{y}}_{t-1}-\hat{\phi_2}\hat{\tilde{y}}_{t-2}}{\hat{\omega}}\right\}$, where $ \hat{\tilde{y}}_t=y_t-\hat{\xi}-\bm{x}_t\trans\hat{\bm{\beta}}$. These plots confirm the validity of using a TGH-AR(2)-e model in which $\epsilon_t \overset{i.i.d.}{\sim}  \mathcal{N} (0, 1)$.

Next, we compare the forecast performance from our TGH-AR-e model to those of the Gaussian AR model for this dataset. We make one-step-ahead forecasts for the whole year of 2014, with a rolling window length of 30 days ($n=720$) for parameter estimation to account for the non-stationarity caused by seasonal effect. To estimate the parameters for the Gaussian AR model, we first remove the diurnal pattern using linear regression with the same harmonics as in the TGH-AR-e model. Then, we find the MLE of the autoregressive parameters from the residuals. 

\begin{figure}[t!] 
	\centering
	\includegraphics[width=0.9\textwidth]{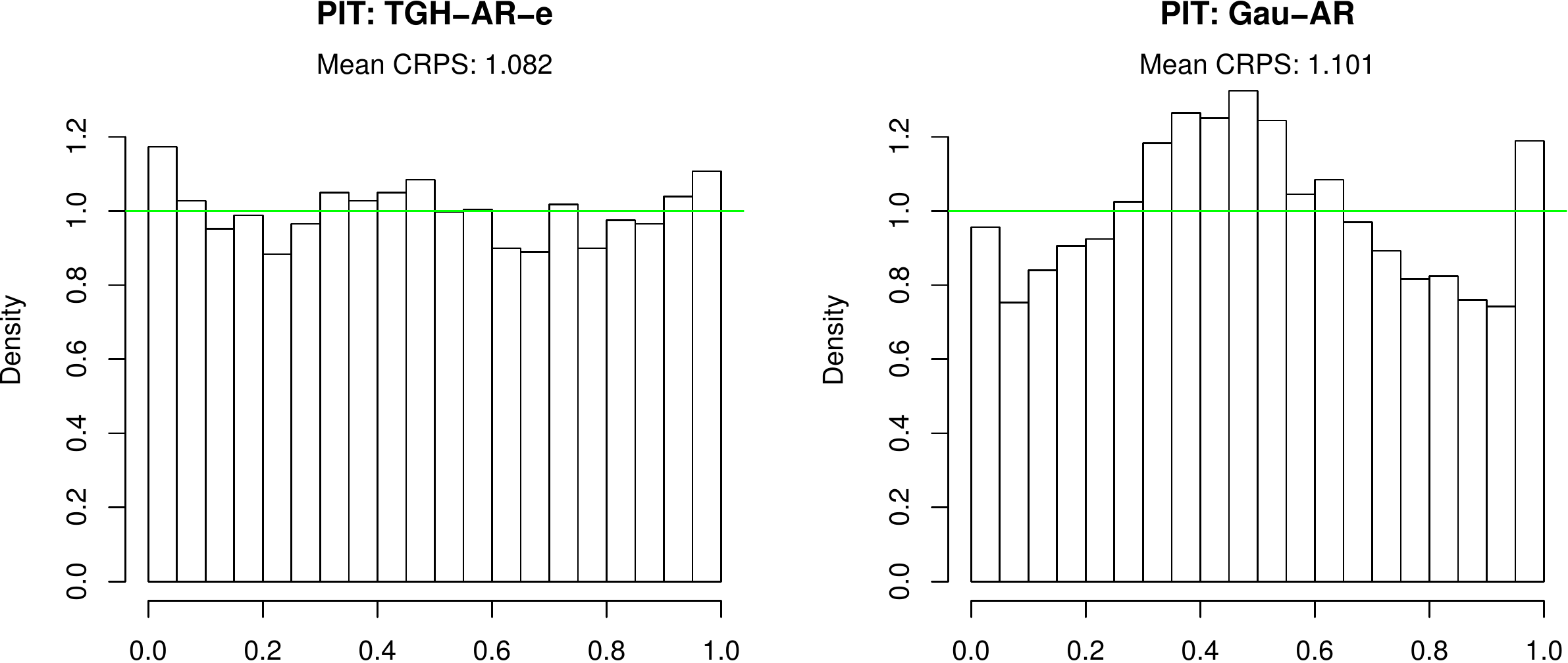}	
	\caption{Histograms of the PIT values of probabilistic forecasts by the TGH-AR-e and Gaussian AR model for one-hour-ahead forecasts over the whole year of 2014.}
	\label{wind2}
\end{figure}

The MAE of the conditional median from the TGH-AR-e model is 4.05; the Gaussian AR forecast has an MAE of 4.28. The RMSEs for the conditional means of the TGH-AR-e model and the Gaussian AR model are 1.47 and 1.50, respectively. The empirical coverage of the 95\% minimum-length CI is 94.2\% for the TGH-AR-e model and 93.3\% for the Gaussian model. We conclude that the point forecasts and CIs based on the TGH-AR-e model are better than those based on the Gaussian AR model for these hourly wind speed data. 
However, it is difficult to see the differences between the forecast results from only these numbers. Figure~\ref{wind2} shows histograms of the PIT values from forecasts based on the TGH-AR-e and Gaussian AR models. By comparing these histograms we see evidently the superiority of fitting the wind speed using a non-Gaussian TGH error in the AR model rather than a Gaussian error.

\section{Discussion}\label{sec:sum}
In this paper, we applied the TGH transformation in a time series context and built two flexible non-Gaussian autoregressive models. The TGH-AR-t model assumes a latent Gaussian process, which is transformed as whole while the TGH-AR-e model transforms the white-noise error term in an autoregressive regression model. The intrinsic difference between the two models can be explained by different data generating mechanisms. For a given time series, a plot like Figure~\ref{rela12} can best help in deciding which model is more suitable. 
We described an efficient parameter estimation procedure for our two models that approximates the maximum likelihood estimator and an order selection procedure based on the approximated likelihood. We derived formulas for point and probabilistic forecast using the two models. We illustrated the empirical performances of estimation, order selection and forecasting of our models through a simulation study. We found that estimating all parameters at once by our estimation procedure, the performance drastically improved compared to sequential estimation that ignores the temporal dependence. Another finding was that the two models yielded different forecasts when calibrated on the same sample, hence proving that the two models are not interchangeable. Simulations also suggested a sample size no less than 250 for a satisfying performance of our models. Finally, we demonstrated the usefulness of our models by applying them to two wind speed datasets at different temporal resolutions. Our TGH-AR models provided a fast simulation method that could emulate wind speed outcome of a gridded climate model and produced competitive forecast results for an observational hourly wind speed dataset from a meteorological station.

The AR models considered in this paper cannot incorporate measurement errors. Extensions of the TGH-AR models to ARMA or state-space models need further research. Also, we feel that an exhaustive comparisons of the many existing transformations, including the Box-Cox, TGH and Sinh-arcsinh transformations \citep{JoPew09}, would be welcome. In a future study, the parameters $g$ and $h$ should be allowed to change smoothly across time, either by imposing a parametric function of $t$ for $g$ and $h$ or by a penalty of smoothness, instead of the moving window scheme we used here. Extension of the TGH framework to a spatio-temporal setting is also promising. 

Additional information and supporting material for this article is available online at the journal's website.

\baselineskip =16pt
\bibliographystyle{apalike}

\end{document}